%% file: manuscript.tex
\documentclass[aps, prb, twocolumn, reprint, superscriptaddress, floatfix, citeautoscript, nofootinbib, longbibliography]{revtex4-2}

\input{preamble/packages.tex}

\input{preamble/abbreviations.tex}

\begin{document} 

\title{Ion-Implanted \eli\ Nuclear Magnetic Resonance in Highly Oriented Pyrolytic Graphite}

\author{John O. Ticknor}
\affiliation{Department of Chemistry, University of British Columbia, Vancouver, BC V6T~1Z1, Canada}
\affiliation{Stewart Blusson Quantum Matter Institute, University of British Columbia, Vancouver, BC V6T~1Z4, Canada}

\author{Jonah R. Adelman}
\altaffiliation[Current address: ]{Department of Chemistry, University of California, Berkeley, CA 94720, USA}
\affiliation{Department of Chemistry, University of British Columbia, Vancouver, BC V6T~1Z1, Canada}

\author{Aris Chatzichristos}
\altaffiliation[Current address: ]{Accenture, Athens 145 64, Greece}
\affiliation{Stewart Blusson Quantum Matter Institute, University of British Columbia, Vancouver, BC V6T~1Z4, Canada}
\affiliation{Department of Physics, University of British Columbia, Vancouver, BC V6T~1Z1, Canada}

\author{Martin H. Dehn}
\altaffiliation[Current address: ]{D-Wave Systems, Burnaby, BC V5G~4M9, Canada}
\affiliation{Stewart Blusson Quantum Matter Institute, University of British Columbia, Vancouver, BC V6T~1Z4, Canada}
\affiliation{Department of Physics, University of British Columbia, Vancouver, BC V6T~1Z1, Canada}

\author{Luca Egoriti}
\affiliation{Department of Chemistry, University of British Columbia, Vancouver, BC V6T~1Z1, Canada}
\affiliation{TRIUMF, Vancouver, BC V6T~2A3, Canada}

\author{Derek Fujimoto}
\altaffiliation[Current address: ]{TRIUMF, Vancouver, BC V6T~2A3, Canada}
\affiliation{Stewart Blusson Quantum Matter Institute, University of British Columbia, Vancouver, BC V6T~1Z4, Canada}
\affiliation{Department of Physics, University of British Columbia, Vancouver, BC V6T~1Z1, Canada}

\author{Victoria L. Karner}
\altaffiliation[Current address: ]{TRIUMF, Vancouver, BC V6T~2A3, Canada}
\affiliation{Department of Chemistry, University of British Columbia, Vancouver, BC V6T~1Z1, Canada}
\affiliation{Stewart Blusson Quantum Matter Institute, University of British Columbia, Vancouver, BC V6T~1Z4, Canada}

\author{Robert F. Kiefl}
\affiliation{Stewart Blusson Quantum Matter Institute, University of British Columbia, Vancouver, BC V6T~1Z4, Canada}
\affiliation{Department of Physics, University of British Columbia, Vancouver, BC V6T~1Z1, Canada}
\affiliation{TRIUMF, Vancouver, BC V6T~2A3, Canada}

\author{C. D. Philip Levy}
\affiliation{TRIUMF, Vancouver, BC V6T~2A3, Canada}

\author{Ruohong Li}
\affiliation{TRIUMF, Vancouver, BC V6T~2A3, Canada}

\author{Ryan M. L. McFadden}
\altaffiliation[Current address: ]{TRIUMF, Vancouver, BC V6T~2A3, Canada}
\affiliation{Department of Chemistry, University of British Columbia, Vancouver, BC V6T~1Z1, Canada}
\affiliation{Stewart Blusson Quantum Matter Institute, University of British Columbia, Vancouver, BC V6T~1Z4, Canada}

\author{Mohamed Oudah}
\affiliation{Stewart Blusson Quantum Matter Institute, University of British Columbia, Vancouver, BC V6T~1Z4, Canada}

\author{Gerald D. Morris}
\affiliation{TRIUMF, Vancouver, BC V6T~2A3, Canada}

\author{Monika Stachura}
\affiliation{TRIUMF, Vancouver, BC V6T~2A3, Canada}
\affiliation{Department of Chemistry, Simon Fraser University, Burnaby, BC V5A 1S6, Canada}

\author{Edward Thoeng}
\affiliation{Department of Physics, University of British Columbia, Vancouver, BC V6T~1Z1, Canada}
\affiliation{TRIUMF, Vancouver, BC V6T~2A3, Canada}

\author{W. Andrew MacFarlane}
\email[Email:]{wam@chem.ubc.ca}
\affiliation{Department of Chemistry, University of British Columbia, Vancouver, BC V6T~1Z1, Canada}
\affiliation{Stewart Blusson Quantum Matter Institute, University of British Columbia, Vancouver, BC V6T~1Z4, Canada}
\affiliation{TRIUMF, Vancouver, BC V6T~2A3, Canada}

\date{\today}

\newcommand{\latin}[1]{\emph{#1}}

\begin{abstract}
We report $\beta$-detected nuclear magnetic resonance of ultra-dilute \elip\ implanted in highly oriented pyrolytic graphite (HOPG). The absence of motional narrowing and diffusional spin-lattice relaxation implies Li$^+$ is not appreciably mobile up to 400 K, in sharp contrast to the highly lithiated stage compounds. However, the relaxation is remarkably fast and persists down to cryogenic temperatures. Ruling out extrinsic paramagnetic impurities and intrinsic ferromagnetism, we conclude the relaxation is due to paramagnetic centers correlated with implantation. While the resulting effects are not consistent with a Kondo impurity, they also differ from free paramagnetic centers, and we suggest that a resonant scattering approach may account for much of the observed phenomenology. 
\end{abstract}

\maketitle

\section{Introduction \label{sec:introduction}}

Graphite is a well-known host for intercalated atomic and small molecular species \cite{Dresselhaus2002}. Among these, lithium is particularly important, as graphitic carbon is commonly the anode in a lithium-ion battery. While a great deal is known about the highly lithiated ordered stoichiometric stage compounds (\eg \ce{LiC6} and \ce{LiC12}), there remain substantial gaps in our understanding of the structure and dynamics of \ce{Li} in graphite, particularly in the dilute limit. In terms of the lithium-carbon phase diagram \cite{Dahn1991}, for $x$ in Li$_x$C$_6$ up to $\sim$5\%, we obtain the $1'$ phase, where graphite maintains its AB (Bernal) stacking with neighboring graphene sheets offset, so that a carbon atom is aligned with the center of an adjacent layer's carbon hexagon. At higher \ce{Li} content (including the stage compounds), the stacking changes to aligned AA with \ce{Li} at hexagonal sites, forming dense layers, which, for the higher stage compounds, are interleaved with unoccupied layers. In contrast, \ce{Li} is thought to be distributed randomly as a dilute solid solution in the $1'$ phase; however, recent measurements suggest its occupancy may still be modulated along the crystallographic $c$-axis \cite{Matsunaga2019}. The occurrence of stage compounds and modulated structures demonstrates a significant Li-Li interaction and complicates determination of the behavior of isolated \ce{Li}. While such interactions are crucial at concentrations relevant for batteries, the simpler case of isolated \ce{Li} provides an important benchmark for theory.

One key question is: how mobile is \ce{Li}? Despite its technological importance, there is a very wide range of reported diffusion coefficients ($D_{\mathrm{Li}}$) \cite{2010-Persson-JPCL-1-1176}. While most of these studies pertain to much higher concentrations, dilute limit macroscopic release measurements find a surprisingly small $D_{\mathrm{Li}}$ \cite{1989-Jungblut-PRB-40-10810}. On the other hand, adsorbed \ce{Li} is highly mobile on the graphite surface \cite{Mandeltort2012}. In the dilute limit, we expect \ce{Li} occupies an interstitial site in the van der Waals gap between adjacent graphene layers. Its mobility is then determined by the energy barrier for hopping to an equivalent near neighbor site (\ie interstitial diffusion). The barrier depends on the interlayer spacing \cite{2011-Ohba-CMES-75-247,2012-Xu-JAP-111-124325}, which is smaller in pure graphite compared to the stage compounds. However, the precise crystallographic site of \ce{Li} in the dilute limit is not known. The analogue of the surface site, the hexagonal site of the stage compounds, may be frustrated by AB stacking that places a carbon atom from the adjacent plane too close, possibly stabilizing an off-center site or forcing the \ce{Li} onto a different high symmetry site.

In addition to structural effects, electronic localization in the adjacent layers may influence the diffusion barrier. Calculations are often done in the concentrated Li regime \cite{2008-Toyoura-PRB-78-214303,2014-Thinius-JPCC-118-2273,2015-Wang-RSCA-5-15985}, where the material is \textit{metallic}. In this case, there appears to be complete charge transfer of the Li valence electron to the graphite band, and the Coulomb potential of interstitial \lip\ is well-screened. Long-standing interest in the screening properties of graphite \cite{DiVincenzo1984,Miesenbock1990} was recently rekindled with the advent of single layer graphene \cite{Fogler2007,Pereira2007,Biswas2007}. The electronic screening cloud in the adjacent layer is generally found to be substantially delocalized. On the other hand, with its relatively high ionization energy, \ce{Li} may not comply with the na\"ive expectation of complete charge transfer in the semimetallic dilute limit, and its binding may be partly covalent \cite{ConardLauginie2002}. This is also suggested by calculations of the adsorption energy \cite{Garay2012,Chan2008} that show a substantial contribution from the transferred charge which remains fairly localized. In particular, as the \ce{Li} concentration decreases, its binding energy increases \cite{Garay2012}.

Aside from its mobility, the electronic structure of dilute \ce{Li} in graphite is interesting in its own right, as alkalis are often used as electron donors to modify the electronic properties of graphite and graphene. The prevailing view, developed over many decades, is that graphite is a semimetal with a carrier density $n \sim$10$^{19}$ \cite{Dillon1978}. Ideally, it is fully compensated with equal numbers of electrons and holes, $n=n_h=n_e$. While $c$-axis dispersion eliminates the perfect Dirac cones, its electronic structure remains closely related to graphene \cite{2006-Partoens-PRB}. The Fermi level cuts through bands composed of carbon $\pi$ orbitals near their energetic extrema along the corners of the Brillouin zone. The resulting Fermi energy is very small ($\sim$30 meV) and the Fermi level lies at the minimum of a V-shaped electronic density of states $\rho(E)$ \cite{Holzwarth1978}. As a consequence, the orbital diamagnetism is very large \cite{Stamenov2005,Raoux2015,Ogata2017}. It is also temperature dependent, increasing in magnitude by $\sim$30\% below 300 K. $n$ is also substantially $T$-dependent, decreasing by $\sim$3-fold from 300 K down to low temperature. The electronic properties generally show significant temperature dependence as degeneracy sets in at low temperature ($T \ll T_F$). For example, in this regime, electron transport exhibits a remarkably strong magnetic field dependence \cite{2005-Du-PRL-94-166601} developing into quantum oscillations at low temperature \cite{2009-Schneider-PRL}. While this conventional view of the electronic properties is supported by many experiments \cite{2008-Kuzmenko-PRL,2012-Schneider-PRL} and detailed calculations \cite{2006-Partoens-PRB}, there remains some controversy (\eg Ref. \onlinecite{2012-Garcia-NJP}), largely due to the possible role of structural imperfections in real samples.

Like a semiconductor, graphite is very sensitive to doping, by either extrinsic species (\eg substitutional \ce{B} acceptors or interstitial \ce{Li} donors) or intrinsic defects, that can cause measurable effects at concentrations as low as 100 ppm \cite{Spain1977}. Doping can be understood in a rigid band picture, but even for the alkalis, the transferred charge per intercalated atom is a parameter that must be determined experimentally \cite{1977-Dresselhaus-PRB-15-3180}.

Here we present nuclear magnetic resonance (\nmr) data for isolated implanted \elip\ in graphite. In principle, \nmr\ is sensitive to the electronic properties of the host, depending on the coupling between the conduction band and the nucleus which, in this case, depends on the extent of hybridization of the \ce{Li} $2s$ orbital with the adjacent carbon $\pi$ states. This is demonstrated, for example, by \ce{^7Li} \nmr\ of the metallic stage compounds, where the resonance is displaced by a small Knight shift (due to the Pauli spin susceptibility) and a $T$-linear Korringa spin-lattice relaxation (\slr) rate that dominates at low temperature \cite{ConardLauginie2002}. At higher temperature, Li diffusion causes motional narrowing \cite{Conard1977} and additional diffusive relaxation in the form of a Bloembergen-Purcell-Pound (BPP) peak around 400 K \cite{1980-Estrade-PB-99-531,1995-Schirmer-ZN-50-643,2013-Langer-PRB-88-094304}, marking the temperature where the elementary hop rate matches the NMR frequencey (in the MHz range) \cite{1948-Bloembergen-PR-73-679}. While there are no previous reports of dilute limit \ce{Li} NMR in graphite, $^{13}$C in pure graphite reveals a substantial orbital (chemical) shift and extremely slow \slr\ rates $\sim$0.002 s$^{-1}$ at 300 K \cite{1988-Hiroyama-65-617}. In contrast, for the dilute implanted muon ($\mu^+$), there is a large and strongly temperature dependent shift and much faster relaxation, features attributed to local moment formation induced by the muon \cite{Jacques2002}.

Aside from orbital diamagnetism, remarkably, some graphite samples also exhibit a permanent magnetic moment, a form of {\it ferromagnetism} \cite{Esquinazi2013,Coey2004}. This is surprising since carbon is usually closed shell, and ideal graphite is no exception. The magnetism appears to be related to structural imperfections, such as point defects \cite{2009-Yang-C-47-1399}, zigzag edges \cite{2009-Cervenka-NP-5-840}, and other grain boundaries \cite{2012-Miao-C-50-1614}, where it cannot achieve a filled shell. The resulting magnetic state is thus inhomogeneous at the atomic scale, and this would be reflected in the NMR as a distribution of internal magnetic fields depending on the distance between the probe nucleus and unpaired electron spins at nearby  defects. Importantly, the ferromagnetic signal has been shown to be eliminated by annealing \cite{2012-Miao-C-50-1614}.

The electronic and magnetic properties of graphite are also sensitive to the type of sample and its purity. Here, we study highly oriented pyrolytic graphite (\hopg) produced by high temperature decomposition of a simple gaseous hydrocarbon followed by vacuum annealing at high temperature and pressure \cite{Moore1973}. With care, the resulting graphite can be very pure, but it is a highly oriented \textit{polycrystal}. The flat crystallites have a very narrow distribution of the orientation of their $c$ axes, while the in-plane directions are completely random.  Though it is composed of micron scale well-oriented crystallites \cite{Ohler2000,2012-Sepioni}, weak interlayer binding makes it prone to turbostratic disorder \cite{Biscoe1942} as well as faults in the AB stacking sequence. For ZYA grade \hopg\footnote{This nomenclature for \hopg\ is based on the mosaic spread of the crystallite orientations and dates from the initial commercial production by Union Carbide Corporation.}, the fraction of ABC stacked rhombohedral graphite \cite{Lipson1942} is negligible \cite{Auslender2021}. At a turbostratic rotational stacking fault (\stf), the adjacent graphene layers are rotated by a random angle about $c$ with respect to the ideal AB stacking. At the surface, this is visible in scanning tunneling microscopy (\stm) \cite{Pong2005} and the resulting Moir\'e fringes have significant consequences for the electronic structure in few-layer graphene \cite{Luican2011}, including the formation of periodic superstructures with an associated strain field \cite{Kazmierczak2021}. Experimentally, less is known about the properties of such defects away from the surface. They probably determine the $c$-axis conductivity \cite{Koren2014}, and there is some evidence that faulted interlayers are inaccessible to \ce{Li} intercalation \cite{Zheng1995}. In contrast, the implanted \elip\ will randomly sample interlayers independent of their stacking character. Interlayers at the \stf\ are still expected to be dilute, but it is unclear how far the fault effects propagate into the adjacent crystallites. \stm\ can detect a buried fault at least several layers deep \cite{Osing1998,Wong2010}, consistent with calculations for sequence stacking faults \cite{Taut2013}. For the graphite used here, the grain size is typically 10s of $\upmu$m in the basal plane \cite{2020-Chatterjee-CM}, with order along the $c$-axis interrupted by stacking faults typically separated by 10s to 100s of nm \cite{Auslender2021,Koren2014}.

We use $\beta$-detected nuclear magnetic resonance (\bnmr) \cite{2015-MacFarlane-SSNMR-68-1,2022-MacFarlane-ZPC-236-757} to measure the temperature and field dependence of both the \slr\ and resonance spectrum for dilute \elip\ implanted into graphite. Importantly, we find neither a diffusion induced $1/T_1$ maximum nor motional narrowing of the resonance line, revealing a suppression of the Li diffusion compared to the stage compounds (\eg \ce{LiC6} and \ce{LiC_{12}}). Overall, the \slr\ is surprisingly fast, weakly field dependent, and well accounted for using a biexponential. It also exhibits a strong temperature dependence between 4 and 100 K, where we find a thermally activated rise of the relative fraction of the fast relaxation given by an activation energy $E_a$ $\approx$ 18 meV. We attribute the fast relaxation to the Li residing near a paramagnetic center, and its phenomenology is discussed.

\section{Experiment \label{sec:experiment}}

An 8 $\times$ 10 $\times$ 1 mm sample of \hopg\ (ZYA) characterized by a mosaic spread of 0.4 $\pm$ 0.1$^{\circ}$ was obtained from TipsNano (Tallinn, Estonia). Elemental analysis performed by HuK Umweltlabor GmbH found impurity concentrations $< 100$ ppb\footnote{A single exception was Ca detected at 240 ppb.}. To ensure a clean surface, the top layers were cleaved with Scotch tape. We adapted an annealing procedure introduced by Miao et al. \cite{2012-Miao-C-50-1614} to eliminate any ferromagnetism. We also studied two additional non-annealed graphite samples: 10 $\times$ 10 $\times$ 2 mm \hopg\ (ZYB) from NT-MDT (Moscow). Using a diamond saw, a sample with perpendicular orientation was cut from one of the samples used in Ref. \onlinecite{Jacques2002} to produce a 1 mm thick slice about 8 $\times$ 4 mm. See the Supplemental Material \cite{suppmater} for details of additional Raman, X-ray, and magnetic characterization of the sample. 

The \eli\ ion-implanted \bnmr\ experiments were conducted at TRIUMF's Isotope Separator and Accelerator facility. The short-lived \eli\ has a nuclear spin $I = 2$, a half-life of 848 ms, a gyromagnetic ratio $\gamma / (2 \pi) =$ 6.3015 MHz T$^{-1}$, and quadrupole moment of $+$32.6 mb \cite{2011-Voss,2015-MacFarlane-SSNMR-68-1}. \elip\ is transported as a low energy (20 keV) ion beam through an electrostatic beamline. The nuclear spin is optically polarized in-flight in a three step process: neutralization in a \ce{Rb} vapor cell, optical pumping with circularly polarized light, and re-ionization in a windowless He gas cell\footnote{The neutral beam is re-ionized with $\sim$60\% efficiency.}. The polarized \elip\ is then delivered to either of the two end-station spectrometers. The mean implantation depth is estimated via SRIM \cite{2010-Ziegler-NIMPRB} simulations to be $\sim$114 nm. The total fluence over the entire experiment is $\sim$3$\times$10$^{13}$ ions cm$^{-2}$, much lower than typical ion irradiation studies. While each \lip\ is estimated to produce $\sim$80 vacancies, the data did not evolve with increasing fluence, confirming that the effects are not caused by cumulative damage. This, however, does not rule out the influence of correlated damage caused by the implanting \lip\ itself, particularly damage at the end of its track that is nearest to the stopping site.

The high field spectrometer uses a high homogeneity 9 T superconducting solenoid, where the applied field $B_0$ is parallel to the incoming beam and $c$-axis. The low field spectrometer involves a far weaker $B_0 = 10$ mT perpendicular to the beam and $c$-axis. In an ultrahigh vacuum chamber ($<10^{-9}$ Torr), the sample is mounted on a He flow coldfinger cryostat surrounded by an \rf\ coil transverse to $B_0$ that is designed to admit the beam from the side. In the experiment, the \elip\ is implanted and the experimental $\beta$-decay asymmetry $A(t)$ is monitored using surrounding scintillation detectors. $A(t)$ is defined by \Cref{eqn_asym}, where $N_{F}(t)$ and $N_{B}(t)$ are histograms of the $\beta$-electron counts as a function of time in a pair of detectors on opposite sides of the sample, here the forward $(F)$ and backward $(B)$. 

\begin{equation}
A(t) = \frac{N_{F}(t) - N_{B}(t)}{N_{F}(t)+N_{B}(t)}
\label{eqn_asym}
\end{equation}

\noindent The parity violating $\beta$-decay (\ce{^8Li} $\to$ \ce{^8Be} $+$ $\bar{\nu}_{e}$ $+$ e$^{-}$) then correlates the spin polarization $p_z(t)$ to the experimental asymmetry $A(t)$ at the time of decay, $A(t) = A_0 p{_z}(t)$. The proportionality constant $A_0$ depends on the field $B_0$, properties of the detectors, and the $\beta$-decay matrix elements. 

Three types of \bnmr\ measurements were taken: 1) spin-lattice relaxation (\slr); 2) resonance, and 3) a quadrupolar resonance comb. In 1), the \elip\ is pulsed (4 s pulse) using a fast electrostatic kicker, followed by a (12 s) beam-off period. This cycle repeats, and the time-resolved $\beta$-decay counts of each iteration are combined to obtain higher statistics, for a typical total run time of $\sim$30 min. In 2), the \elip\ is continuously implanted while the \rf\ is stepped slowly through a range of frequencies around the nuclear Larmor frequency $\omega_0 = 2\pi \nu_0 = \gamma B_0$ using a frequency step that is a small fraction of the linewidth. At each step, the $\beta$ counts are accumulated for an integration time (typically 1 second). The scan is then repeated alternating both the helicity of the laser light (\ie left/right sense of circular polarization) and the frequency sweep direction to minimize systematics and obtain higher statistics. The scans are then combined to yield the measured spectrum. When the RF frequency matches $\nu_0$, it causes rapid spin precession about the direction of the RF field, and the asymmetry is reduced. To assess quadrupolar effects, it is useful to combine the two helicities separately, since they should contain opposite quadrupolar satellites. Examples of these ``helicity resolved'' spectra are shown in \Cref{apx:split_hel}. In 3), the \rf\ is an equal amplitude sum of 4 frequencies: $\tilde{\nu}_0 \pm \tilde{\nu}_q$ and $\tilde{\nu}_0 \pm 3\tilde{\nu}_q$. The fixed parameter $\tilde{\nu}_0$ is chosen as the center of the resonance from 2). For a resonance split into the $2I = 4$ quadrupolar satellites, this comb can simultaneously saturate all the single quantum transitions when the stepped parameter $\tilde{\nu}_q$ matches the quadrupole frequency $\nu_q$, strongly enhancing the resonance amplitude. As before, the resonance condition is marked by a pronounced reduction of the asymmetry. The comb measurements also alternate helicity and frequency step direction. See Ref. \onlinecite{1993-Minamisono-HI-80-1315} for further details.

\clearpage

\section{Results\label{sec:results}}

\subsection{Spin-Lattice Relaxation\label{subsec:SLR}}

\begin{figure}[H]
\centering
\includegraphics[height=3in]{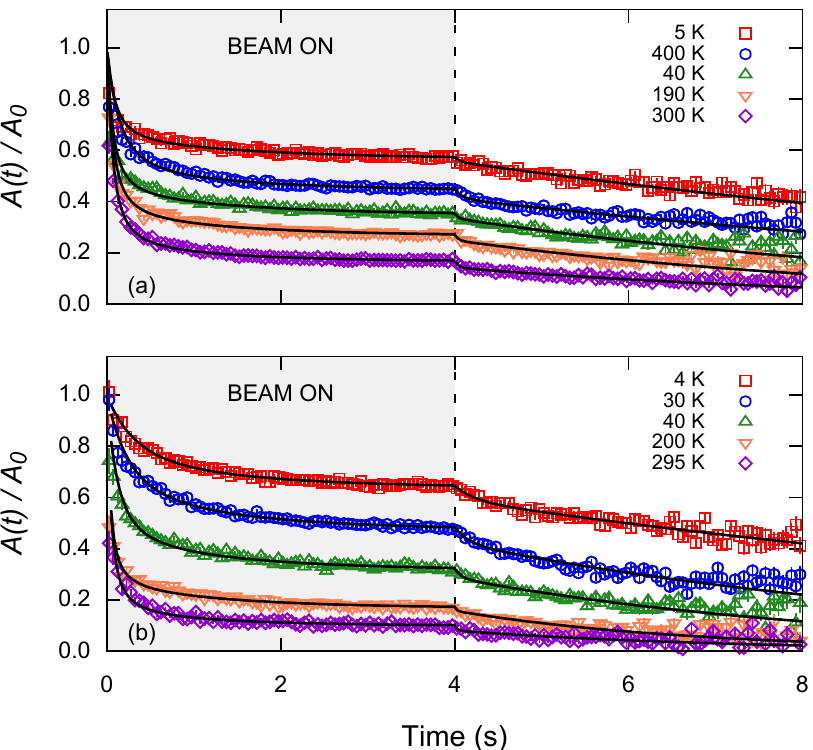}
\caption{Selected \slr\ data [$A(t)$ from \Cref{eqn_asym}] for $B_0 =$ 6.55 T $\parallel$ $c$-axis (a) and 10 mT $\perp$ $c$-axis (b) for \elip\ implanted in \hopg\ (ZYA). During the 4 second beam pulse, the asymmetry relaxes to a dynamic equilibrium, and to thermal equilibrium after. The error bars increase with time due to the lifetime of \eli. The data, represented as histograms, have been binned by a factor of 5 for clarity.}
\label{fig:HOPG_SLR_Spectra_Fits}
\end{figure}

Representative \eli\ \slr\  measurements are shown in \Cref{fig:HOPG_SLR_Spectra_Fits}. During the beam pulse the asymmetry relaxes to a dynamic equilibrium value, while after, it relaxes to its thermal equilibrium value near zero, giving the \bnmr\ ``recovery'' curve its characteristic bipartite shape. We find that a biexponential relaxation function is the simplest model that describes the data well. Specifically, at time $t$ after its implantation the \eli\ polarization follows \Cref{eqn_bi_exp}, where $\lambda_{\textrm{slow}} = 1 / T_1^{\textrm{slow}}$, $\lambda_{\textrm{fast}} = 1 / T_1^{\textrm{fast}}$, and $f_{\textrm{slow}}+f_{\textrm{fast}}=1$.

\begin{equation}
p_z(t) = f_{\textrm{slow}}e^{-\lambda_{\textrm{slow}}t}+f_{\textrm{fast}}e^{-\lambda_{\textrm{fast}}t},
\label{eqn_bi_exp}
\end{equation}

\noindent The data were fit \cite{2021-Fujimoto-JOSS-6-3598} to this function convoluted with the square beam pulse to yield the curves shown in \Cref{fig:HOPG_SLR_Spectra_Fits}.

%% FIGURE %%

%% FIGURE %%

\begin{figure}[H]
\centering
\includegraphics[height=4in]{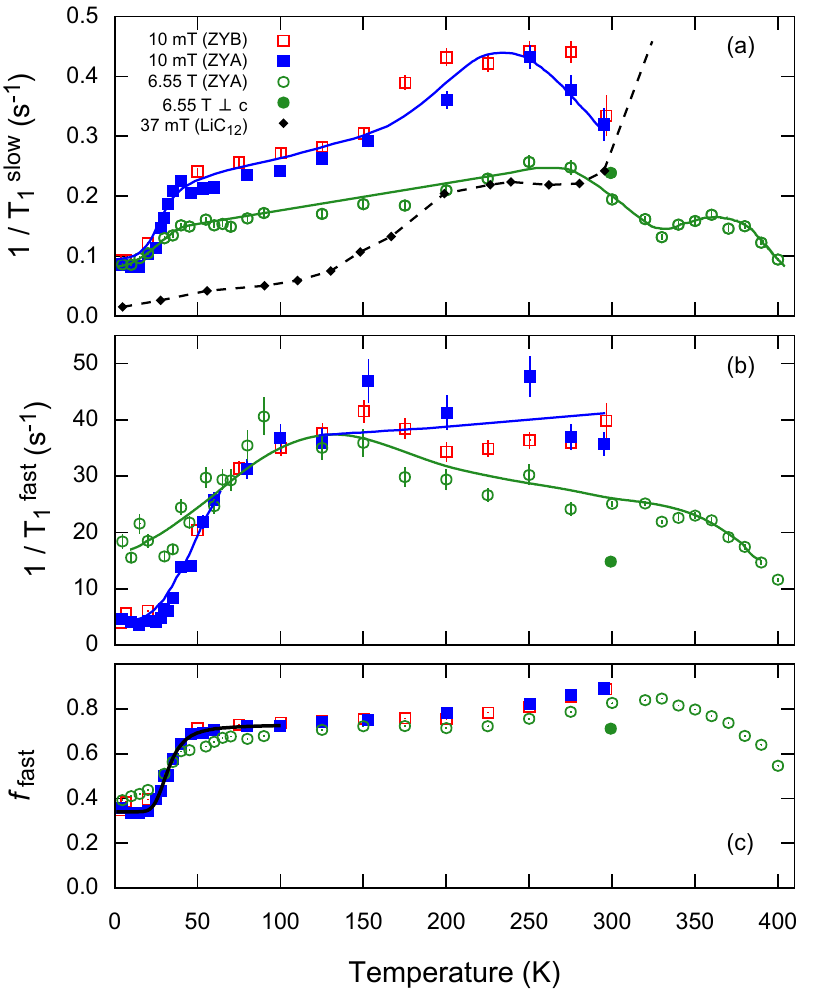}
\caption{Temperature dependence of the slow component \slr\ rate $1/T_1^{\textrm{slow}}$ (a), fast component \slr\ rate $1/T_1^{\textrm{fast}}$ (b), and relative fraction $f_{\textrm{fast}} = 1 - f_{\textrm{slow}}$ (c) for \hopg. For comparison, also shown are \eli\ $1/T_1$ values reported for \ce{LiC12} \cite{1995-Schirmer-ZN-50-643} and a single measurement for $B_0 \perp$ 6.55 T at 300 K. In the biexponential fits for \hopg, all parameters except $A_0$ were allowed to vary freely. Note that the \slr\ in \ce{LiC12} was fit only to a single exponential, that is, without a faster relaxing component. By comparison, in \hopg, the relaxation between $\sim$50-300 K is dominated by the fast component. The solid lines in (a) and (b) serve to guide the eye, while the solid line in (c) is a fit to a sigmoidal function [\Cref{eqn_sigmoid}] described in \Cref{sec:discussion}. The dashed line in (a) follows the \ce{LiC12} data and its offscale exponential rise.}
\label{fig:HOPG_SLR_T1_Slow_Fast_Fraction}
\end{figure}

The initial asymmetry $A_0$ was calibrated in slowly relaxing reference samples: single crystal \ce{MgO} ($B_0 =$ 6.55 T) and \ce{Au} foil ($B_0=$ 10 mT), finding $A_0 = 0.095(3)$ at 6.55 T and 0.094(4) at 10 mT which were used to normalize the data, as shown in \Cref{fig:HOPG_SLR_Spectra_Fits}, and to avoid overparameterization, such that the free parameters are the two rates and the relative fraction (\ie $A_0$ is fixed). At early times, the spectra are dominated by the fast relaxing component, but this contribution vanishes shortly after the pulse, the relaxation is almost entirely the slow component. The overall global fit quality was good at each field: $\chi^2_{\textrm{6.55 T}} = 1.08$ and $\chi^2_{\textrm{10 mT}} = 1.07$, and the resulting parameters are shown in \Cref{fig:HOPG_SLR_T1_Slow_Fast_Fraction}. For a more detailed account of the biexponential fit, see \Cref{apx:SLR_fit_func}.  

%% FIGURE %%

%% FIGURE %%

The most salient feature of the data is the remarkably fast relaxation, which persists even in the limit of low temperature and high field. This is reflected in the $1/T_1^{\textrm{fast}}$ data shown in \Cref{fig:HOPG_SLR_T1_Slow_Fast_Fraction}\textcolor{blue}{b}. Here, only a weak field dependence is observed at the temperature extremes. At low field, $1/T_1^{\textrm{fast}}$ increases monotonically with temperature until it plateaus near 150 K. However, at high field, it passes through a broad maximum over the same temperature range as the plateau. The relative fraction of the fast component $f_{\textrm{fast}}$ (see \Cref{fig:HOPG_SLR_T1_Slow_Fast_Fraction}\textcolor{blue}{c}) predominates, except at the lowest temperatures. On warming from 4 K, it rises monotonically until it passes through a maximum near 300 K and its trend above 300 K appears correlated to its rate $1/T_1^{\textrm{fast}}$.

Inherent to the biexponential is the slower relaxing component $1/T_1^{\textrm{slow}}$ (see \Cref{fig:HOPG_SLR_T1_Slow_Fast_Fraction}\textcolor{blue}{a}), which displays a slightly stronger field dependence, though it converges to about the same value around 10 K. At higher temperatures, we observe a broad maximum in $1/T_1^{\textrm{slow}}$ near 250 K at low field, and a secondary maximum near 360 K at high field. $1/T_1^{\textrm{slow}}$ is also comparable to \ce{LiC_{12}} (\Cref{fig:HOPG_SLR_T1_Slow_Fast_Fraction}\textcolor{blue}{a}), however, the pronounced increase above 250 K, characteristic of fast hopping, is absent in our data. Finally, we find good quantitative agreement between the different samples, indicating that the observed relaxation is intrinsic to \eli\ implanted in graphite and is unaffected by sample purity or mosaic spread.

\subsection{Resonance Spectra\label{subsec:RES}}

\begin{figure}[H]
\centering
\includegraphics[height=3in]{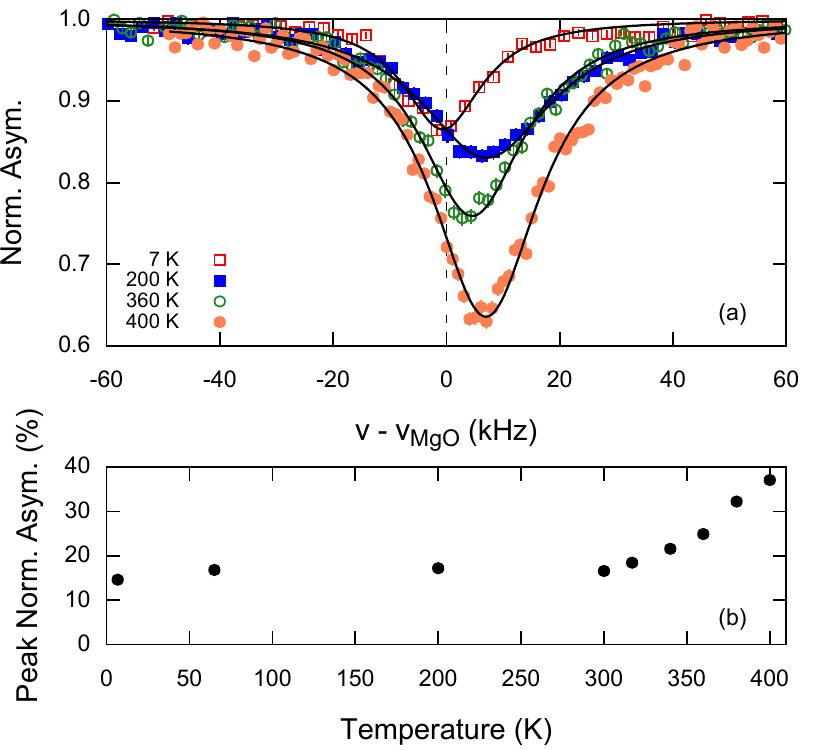}
\caption{Time-integrated resonance spectra (a) for \elip\ implanted in HOPG (ZYA) at $B_0 =$ 6.55 T for several temperatures and the peak amplitude (b) relative to the off-resonance baseline asymmetry. For both, the data are normalized by the off-resonance baseline asymmetry. The data in (a) are plotted relative to the reference frequency $\nu_{\textrm{MgO}}$, measured in a \ce{MgO} (100) single crystal. The solid lines denote Lorentzian fits.}
\label{fig:HOPG_RES_1f_Stacked}
\end{figure}

\begin{figure}[H]
\centering
\includegraphics[height=3.5in]{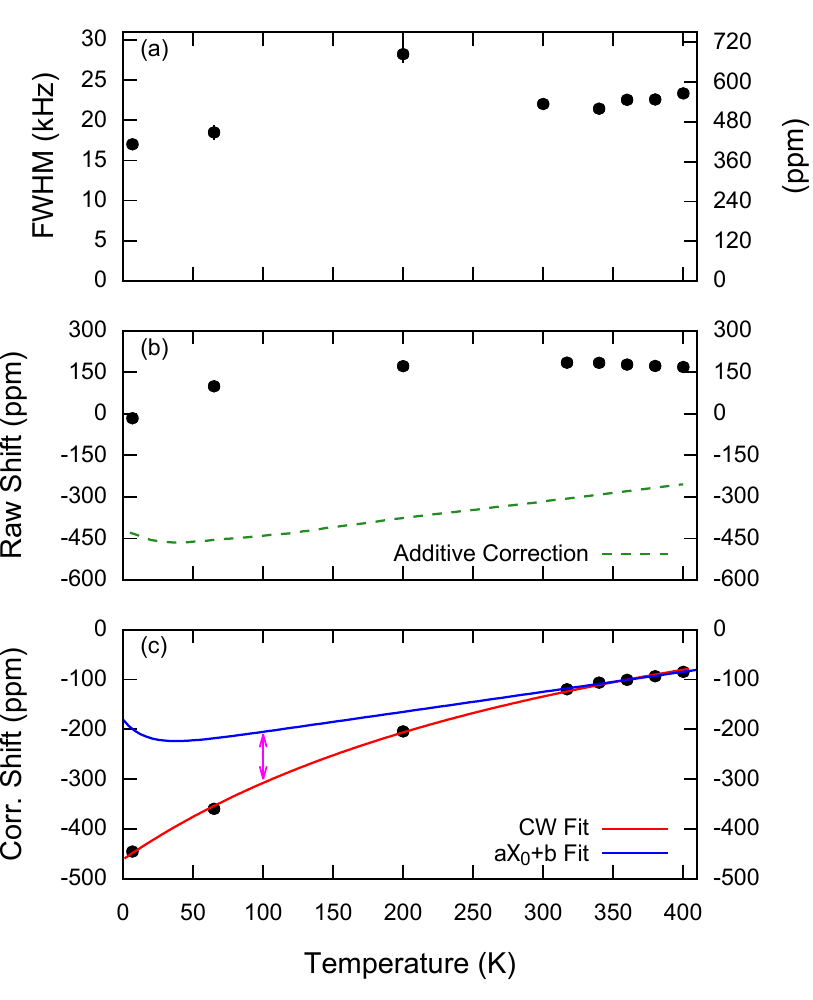}
\caption{The full-width at half-maximum (FWHM) of the Lorentzian fits of the resonances (a), the raw shift (in ppm) relative to $\nu_{\textrm{MgO}}$  (b), and the demagnetization corrected shift (c). The dashed line in (b) shows the additive demagnetization correction (see \Cref{apx:shift}). The solid red line in (c) is a fit to a modified Curie-Weiss relationship [\Cref{eqn_curie}], while the solid blue line shows a fit to the bulk diamagnetic susceptibility $\chi_0(T)$, as described in the text.}
\label{fig:HOPG_RES_width_shift}
\end{figure}

\Cref{fig:HOPG_RES_1f_Stacked} shows selected resonance spectra at 6.55 T as a function of temperature. The lines are broad and lack any fine structure. In particular, no quadrupolar splitting is evident. A single Lorentzian describes the resonance well as shown by the fit lines in \Cref{fig:HOPG_RES_1f_Stacked}. On warming, the position varies by $\sim$10 kHz between 7 and 400 K. The width (see \Cref{fig:HOPG_RES_width_shift}\textcolor{blue}{a}) is nearly temperature independent. More pronounced changes are evident in the amplitude (\Cref{fig:HOPG_RES_1f_Stacked}\textcolor{blue}{b}), which increases significantly above 300 K. Though there is no resolved splitting, when the spectra are decomposed into separate helicities, the peak position differs slightly (see \Cref{apx:split_hel}), consistent with a small unresolved splitting \cite{2015-MacFarlane-SSNMR-68-1}.

%% FIGURE %%

The resonance also does not exhibit the two component character that might be expected from the biexponential SLR. This is, however, not surprising, since the resonance amplitude is determined by the time-average polarization proportional to $(\tau/T_1 + 1)^{-1}$.
This magnifies the resonance amplitude of the slow component about 20-fold over the fast component, and the observed resonance is predominantly that of the slow relaxing fraction.

The raw $(\delta)$ and demagnetization corrected $(\delta_{\textrm{C}})$ NMR shifts are obtained from the Lorentzian fits as detailed in \Cref{apx:shift} and are shown in \Cref{fig:HOPG_RES_width_shift}.
The temperature dependence of $\delta_{\textrm{C}}$ is fit to a Curie-Weiss relationship in \Cref{eqn_curie},
where $a$, $b$, and $\theta$ are free parameters.

\begin{equation}
\delta_{\textrm{C}} = b - \frac{a}{T+\theta}
\label{eqn_curie}
\end{equation}

\noindent The fit yields values of $a=$ 292(53) $\times$ 10$^3$ ppm K, $b=$ 290(46) ppm, and $\theta=$ 389(48) K and is depicted as the solid red line in \Cref{fig:HOPG_RES_width_shift}\textcolor{blue}{c}.

\subsection{Comb Spectra\label{subsec:COMB_RES}}

\begin{figure}[H]
\centering
\includegraphics[height=2.5in]{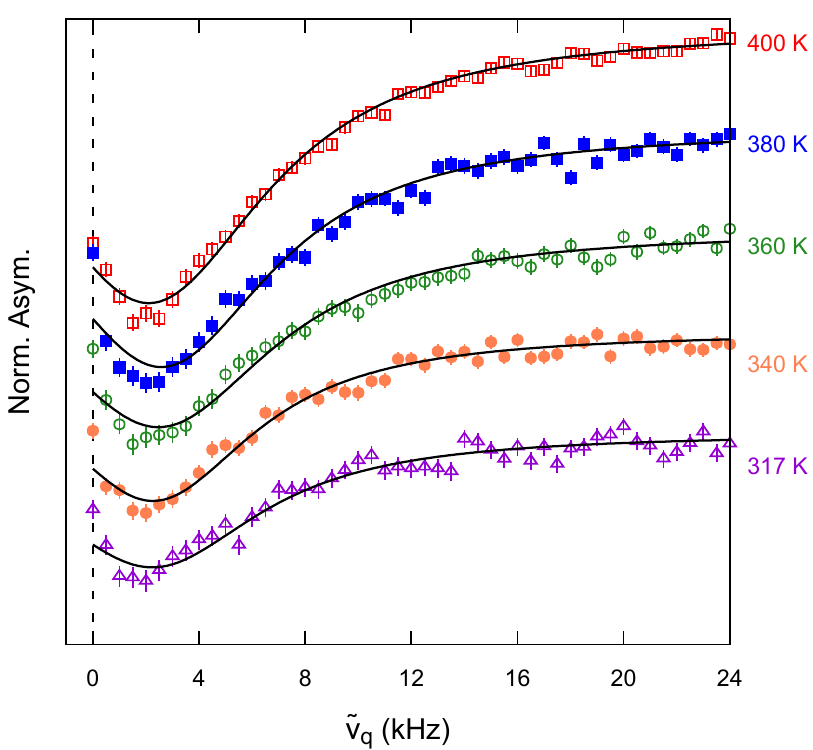}
\caption{Frequency comb spectra at $B_0=$ 6.55 T. $\tilde{\nu}_q$ is the RF comb splitting parameter. At each temperature, the data are normalized to the baseline asymmetry and offset vertically for clarity. The solid lines are single Lorentzian fits, which here only approximate $\nu_q$ due to the absence of clearly resolved $\frac{1}{3}\nu_q$ and $3 \nu_q$ peaks.}
\label{fig:HOPG_RES_1w_Stacked}
\end{figure}

The frequency comb spectra shown in \Cref{fig:HOPG_RES_1w_Stacked} supplement the resonance measurements above. The spectra are quite broad, but their width is about half that of the single tone resonances in \Cref{fig:HOPG_RES_1f_Stacked}\textcolor{blue}{a}. Notably, the resonance is centered at a small, but non-zero, $\tilde{\nu}_q\sim$2 kHz between 317 and 400 K. This implies that, although there appears to be a broad distribution of quadrupole splittings, there is a small non-zero average value. Conventionally, $\nu_q$ is defined in \Cref{eqn_quad} in terms of the product of the principal component of the electric field gradient (EFG) $V_{zz}$ and the nuclear electric quadrupole moment $eQ$. 

\begin{equation}
\nu_q = \frac{3 e Q V_{zz}}{4I(2I-1) h} = \frac{e Q V_{zz}}{8 h}
\label{eqn_quad}    
\end{equation}

\noindent Note that the EFG (and thus quadrupole splitting) is characteristic of Li at a specific crystallographic site. The parameters of the Lorentzian fits shown in \Cref{fig:HOPG_RES_1w_Stacked} are listed in \Cref{tab:HOPG_1w_fit_data}, revealing that the distribution of $\nu_q$ (both average and width) is quite independent of temperature in this range.

%% FIGURE %%

%% FIGURE %%

%% TABLE %%
\begin{table}[H]
\caption{Lorentzian parameters for the comb spectra shown in \Cref{fig:HOPG_RES_1w_Stacked}. Statistical errors are denoted in parentheses.}
\centering
\begin{tabular}{ccc}
\hline
Temperature (K) & $\nu_q$ (kHz) & FWHM (kHz) \\ \hline
317     &        2.2(2)     &        9.9(6)             \\
340     &        2.3(1)     &        9.3(4)             \\
360     &        2.5(1)     &        10.6(4)             \\
380     &        2.5(1)     &        10.2(3)             \\ 
400     &        2.1(1)     &        10.8(2)             \\ \hline
\end{tabular}
\label{tab:HOPG_1w_fit_data}
\end{table}

%% TABLE %%

\section{Discussion \label{sec:discussion}}

Perhaps the most surprising discovery is that the high field \slr\ is notably fast even at low temperatures. It is much faster than $^{13}$C in graphite \cite{1988-Hiroyama-65-617} and more than an order of magnitude faster than \eli\ in semimetallic bismuth \cite{bismuth2014}. This suggests some form of electronic paramagnetism is responsible. However, the purity of the sample, as well as the similarity between samples from different sources, suggest it cannot be due to extrinsic magnetic impurities. Moreover, very dilute magnetic microparticles, found in some \hopg\ \cite{Venkatesan2013}, would cause relaxation only for the \elip\ in their immediate vicinity, giving rise, at most, to a very small amplitude fast relaxing component, inconsistent with the data. To test whether the inhomogeneous structural disorder (\eg stacking faults) or defect-related ferromagnetism \cite{Esquinazi2013,Coey2004} might be responsible, we annealed the ZYA sample to eliminate it \cite{2012-Miao-C-50-1614} and found no substantial difference (see \Cref{apx:anneal}). We now consider possible sources of the fast relaxation.

\subsection{Relaxation from Free Carriers}

The \slr\ is far too fast to be Korringa relaxation from the free carriers. This is demonstrated by comparison with the intercalated stage compounds, where doping has moved the Fermi level $E_F$ up into the conduction band, and $\rho(E_F)$ is $\sim$100$\times$ larger \cite{Holzwarth1978,Gerischer1987}. In \ce{LiC_{12}}, it is evident \cite{1995-Schirmer-ZN-50-643} as the low $T$ slope (reproduced in \Cref{fig:HOPG_SLR_T1_Slow_Fast_Fraction}\textcolor{blue}{a}), while in \ce{LiC6} it is somewhat larger \cite{1980-Estrade-PB-99-531}. In a metal, the slope is proportional to $\rho(E_F)$ squared. Assuming the hyperfine coupling is the same, it should be a factor of $10^4$ smaller in pure graphite, making it immeasurably slow. The coupling will probably not be identical (due to both stacking and interlayer spacing), but it is very unlikely that this could compensate for the vast difference in carrier density. The temperature dependence in \Cref{fig:HOPG_SLR_T1_Slow_Fast_Fraction} is also inconsistent with Korringa relaxation. In a semimetal, the energy dependence of $\rho(E)$ near $E_F$ transforms the linear metallic dependence to a supralinear power law. Additionally, one must account for the small $E_F$ \cite{Hoch2005}, but neither of these effects provides anything qualitatively similar to the observed temperature dependence.

\subsection{Relaxation from \lip\ Diffusion}

Diffusive motion of the implanted \elip\ is also unable to explain the data. In this case, the \slr\ would be predominantly caused by modulation of the quadrupole interaction due to stochastic hopping of \elip\ within the lattice. Again this is demonstrated in the stage compounds, which exhibit the characteristic field-dependent \bpp\ peak at $\sim$400 K \cite{1995-Schirmer-ZN-50-643} (see the dashed line that rises exponentially offscale in \Cref{fig:HOPG_SLR_T1_Slow_Fast_Fraction}\textcolor{blue}{a}). Note that the minor bump evident in the \ce{LiC_{12}} data around 200 K is not due to long-range diffusion. Its origin is unclear, but it coincides with a discontinuity of the measured EFG in \ce{LiC_{6}} \cite{1988-Heitjans-SM-23-257}, and interestingly, we see a similar feature in $1/T_1^{\textrm{slow}}$. Unlike in the Li-dense compounds, however, the low field $1/T_1^{\textrm{slow}}$ here \textit{decreases} between 250 and 300 K. Since diffusive relaxation is an activated process, it should fall exponentially when the temperature falls below the hopping barrier on the low temperature side of the \bpp\ peak. In contrast, our relaxation persists to low temperature showing substantial variation (primarily below 100 K), with relaxation rates that are only weakly field dependent. This is in sharp contrast to the expected field dependence of diffusive relaxation. Furthermore, below the \bpp\ peak, the resonance width should exhibit motional narrowing when the hop rate exceeds the static resonance linewidth, here $\sim$15 kHz. This is also seen in the stage compounds (\eg \ce{LiC6}), where the narrowing is evident at $\sim$280 K \cite{Conard1977,2013-Langer-PRB-88-094304}. As shown in \Cref{fig:HOPG_RES_width_shift}, our width continues to broaden up to 400 K with no sign of narrowing. From these observations, we conclude \textit{the isolated implanted \ce{Li} is not appreciably mobile in \hopg\ up to 400 K}. Though surprising, this is consistent with dilute limit macroscopic release measurements \cite{1989-Jungblut-PRB-40-10810}. Specifically, at 400 K, the hop rate must be substantially less than the linewidth. The reduced mobility (relative to the stage compounds) suggests the barrier is substantially higher. In part, this may be due to the smaller interlayer spacing \cite{2012-Xu-JAP-111-124325}, but reduced charge transfer and partially covalent binding may also play a role.

\subsection{\lip\ Induced Carrier Localization}

Having ruled out several potential mechanisms for the fast relaxation, we now return to electronic paramagnetism. Instead, of the pre-existing unpaired spins, we now consider paramagnetic defects correlated with the \ce{Li} itself. While some atomic intercalates are known to give rise to local moments (\eg \ce{F} \cite{Panich2001}, \ce{H} \cite{Esquinazi2003}, and possibly $\mu^+$ \cite{Jacques2002}), there is no evidence that alkalis do. For example, alkali intercalation certainly alters graphite's intrinsic (diamagnetic) susceptibility $\chi_0$, but it does not introduce a Curie term \cite{DiSalvo1979}. It seems unlikely that the \lip\ potential could give rise to a bound state in semimetallic graphite (see the discussion in Ref. \onlinecite{Yazyev2010}), but if the electronic spectrum is gapped (\eg by disorder or the magnetic field \cite{Luican2014,Niimi2006}), it may become possible. More likely, an unpaired spin may be the result of implantation damage.

\subsection{\lip\ Implantation-Related Magnetic Defects}

As with other properties, the study of radiation defects in graphite has a long history \cite{Telling2007}, in part due to its application as a neutron moderator in nuclear reactors. The most common long-lived damage from low-dose ion beams is the vacancy-interstitial (Frenkel) pair caused when the implanting ion displaces a carbon atom from its normal lattice site. In graphite, this requires $\sim$20 eV \cite{Thrower1978}, so the \elip\ continues some distance before stopping. In some cases, defects of this type are associated with an unpaired electron. Electron spin resonance (\esr) is the natural method to detect and characterize such defects \cite{Corbett1983}. However, in graphite, they do not produce a distinct \esr\ signal. Instead, exchange coupling with the carrier spins (which have their own well-known \esr) yields a composite resonance \cite{Matsubara1991,Kazumata1983,Kazumata1986,2023-Vorobiev-JPCC}. The displaced interstitial carbon is not likely stable as an atomic species but probably forms a bridge between adjacent sheets \cite{Telling2007}. While this does not rule it out as the source of paramagnetism, we focus on the vacancy which has been studied more extensively. In fact, native point magnetic defects (such as the isolated vacancy) may be related to graphite ferromagnetism \cite{Coey2004,2009-Cervenka-NP-5-840,Esquinazi2013}. They are also important in single layer graphene \cite{Yazyev2010}, where they cause spin relaxation of the mobile carriers \cite{Lundeberg2013,Kochan2014,Roche2014,Avsar2020}. The electronic structure of the vacancy has been studied in considerable detail \cite{Zunger1978,Nanda2012}, and \stm\ measurements, while not directly spin sensitive, strongly suggest a paramagnetic state \cite{Ugeda2010}.

The magnetic field of such a local moment would influence the NMR of a nearby nucleus. Its time-average produces a static field that shifts the resonance by $K_{\mathrm{imp}} \propto \chi_{\mathrm{imp}}$, where $\chi_{\mathrm{imp}}$ is the impurity spin susceptibility, which would be Curie-like in the simplest case of an uncoupled moment. Additionally, temporal fluctuations of the moment give rise to field fluctuations at the probe nucleus whose transverse components cause its spin to relax. As shown in \Cref{fig:HOPG_RES_width_shift}\textcolor{blue}{c}, the measured shift does not show the Curie dependence expected from either the \esr\ of irradiated graphite \cite{Kazumata1986} or vacancies in graphene \cite{Nair2013}. Instead, it can be described by a negative Curie-Weiss (CW) law with a large positive offset. The negative value of the shift implies a negative hyperfine coupling between \eli\ and the local moment. If the coupling was a direct hybridization between the localized carbon orbital and the Li $2s$ orbital, one would expect the coupling to be positive. On the other hand, negative \eli\ shifts are also found in dilute magnetic Ga$_{1-x}$Mn$_x$As \cite{Song2011} and Bi$_{2-x}$Mn$_x$Te$_3$ \cite{2020-McFadden-PRB-102-235206}, where the coupling is probably mediated by valence band holes. Superficially, the CW dependence of $K$ resembles a Kondo impurity in a metal, where $\theta=T_K$, the Kondo temperature, as has been  shown by conventional NMR in dilute alloys \cite{Boyce1974,Alloul1977}. Despite the very low $\rho(E_F)$ (and inconsistent with the reported Curie-like response), it has been suggested that the vacancy may act as a Kondo-like defect in graphene \cite{Fritz2013,Jiang2018}. However, our $\theta$ is much larger than the estimated $T_K$, and we conclude it must have a different origin.

In fact, $K(T)$ likely also contains a substantial chemical (orbital) shift $K_{\mathrm{orb}}$ that is temperature dependent, following the bulk diamagnetic $\chi_0(T)$ (\ie $K = K_{\mathrm{imp}}(T) + K_{\mathrm{orb}}(T)$). To separate these contributions, we assume $\chi_{\mathrm{imp}}$ (and hence $K_{\mathrm{imp}}$) is negligible above room temperature, and fit the shift to $A\chi_0(T) + B$ as shown in \Cref{fig:HOPG_RES_width_shift}\textcolor{blue}{c}, where $A$ and $B$ are constants. At low $T$, $K$ diverges from this curve. We attribute the ``excess'' shift (indicated by the arrow in \Cref{fig:HOPG_RES_width_shift}\textcolor{blue}{c}) to the additive $K_{\mathrm{imp}}$. While it appears magnetic, in the sense that it increases with decreasing $T$, it follows neither a Curie nor a CW dependence. This may indicate a problem with the decomposition, or that there is a more complicated  temperature dependent screening of the local moment at play.

We now consider the resonance linewidth. If there are other more distant magnetic vacancy defects, the distribution of distances to the stopped \eli\ should give rise to magnetic {\it broadening} that also scales with $\chi_{\mathrm{imp}}$, growing larger at low $T$. Instead, we find the linewidth to be both large and nearly $T$-independent (\Cref{fig:HOPG_RES_1f_Stacked}\textcolor{blue}{a}). This suggests the density of magnetic defects is sufficiently low that their broadening effect is negligible in comparison to some other source of linewidth. The weak $T$ dependence suggests it is static (with nearly the same value at 4 K and 400 K), and thus reflects some type of microscopic inhomogeneity. It cannot be due to nuclear dipolar broadening, which is negligible as a result of the low abundance of $^{13}$C. Inhomogeneity of the demagnetizing field in the non-ellipsoidal sample is potentially a large effect in graphite \cite{Goze-Bac2002}, but because the implanted \eli\ samples a small region at the center of the front face, defined by the beamspot ($<$2 mm diameter) and average implantation depth ($\sim$114 nm), it should be negligible, provided the sample acts as an effectively monolithic diamagnet. On the other hand, if grain boundaries interrupt the orbital response, the corresponding shift would be inhomogenous, giving rise to a magnetic broadening. However, this would be proportional to $\chi_0(T)$, so the line should \textit{broaden} from 300 K to 50 K, while the data shows the opposite trend with a slight narrowing, and the source of the width must lie elsewhere.

The most likely source of the width is a distribution of quadrupolar splittings. In fact, we do find a small average splitting using the frequency comb (\Cref{fig:HOPG_RES_1w_Stacked}), $\langle \nu_q \rangle \sim$2 kHz, corresponding to an \efg\ of 0.2 $\times 10^{20}$ V m$^{-2}$. This very small value is consistent with time differential perturbed angular distribution measurements of another dilute limit implanted alkali $^{22}$Na \cite{1993-Martin-HI-77-315}.  If the \elip\ site has 3-fold symmetry along the $c$-axis (\ie any hexagonal ``pocket'' site), then the \efg\ will be axial, and there would be no powder broadening in the ideally oriented polycrystal. Thus, the observed broadening would then imply that the \efg\ varies considerably. One possible source for this is turbostratic (and other) stacking faults together with their associated strain fields \cite{Kazmierczak2021}, but these are probably too rare to account for the measured width, unless their effects propagate further from the fault than expected. On the other hand, if \ce{Li} occupies a non-axial site, then the random perpendicular orientation of the crystallites will give rise to powder broadening. It would be interesting to compare the \efg\ with calculations to aid in determining the site. With a better understanding of the site in an ideal graphite single crystal, it might be possible to relate the quadrupolar broadening to a specific model of the disorder (\eg Ref. \onlinecite{Puech2019}).

Finally, we return to the \slr. The fluctuating dipolar field of a nearby localized electron spin will cause the \eli\ spin to relax at a rate given by \Cref{eqn_impurity}, where $\tau_1$ is the \slr\ time for the electron spin (its longitudinal correlation time), $\omega_0$ is the NMR frequency \cite{Alloul1974}, and $D$ is the dipolar hyperfine coupling constant.

\begin{equation}
\frac{1}{T_1} = \gamma^2 D^2 kT \chi_{\mathrm{imp}} \frac{\tau_1}{1+ (\omega_0 \tau_1)^2}
\label{eqn_impurity}
\end{equation}

\noindent From the near independence of the measured $1/T_1$ on applied field, we infer that our measurements must be in the fast-fluctuating regime where $\omega_0 \tau_1 \ll 1$ even at the highest field (and lowest temperature), meaning $\tau_1 \ll 3$ ns. One might expect the \esr\ linewidth to be $1/\tau_1$;  however, the bottleneck effect apparently limits it to a much smaller value \cite{Kazumata1986}.  It has been shown that the \nmr\ $1/T_1$ provides a measure of $\tau_1$ for a magnetic impurity, even when the \esr\ is bottlenecked \cite{Alloul1974}. For a Kondo impurity in a metal, $1/\tau_1$ follows a Korringa temperature dependence down to $T_K$ and then crosses over to a low $T$ constant $k_B T_K/\hbar$. Since we are in the fast-fluctuation regime, and provided $T \chi_{\mathrm{imp}}$ is roughly constant, \Cref{eqn_impurity} shows that $1/T_1 \propto \tau_1(T)$. In sharp contrast to a Kondo defect, $1/T_1$ {\it slows} at low temperatures, reflecting an order of magnitude {\it increase} in $1/\tau_1$ between 100 K and 40 K where it attains a low temperature limit. Below 100 K, the electronic properties of graphite are strongly temperature dependent. For example, the temperature dependent resistivity changes as low energy phonons freeze out, and the band electrons become degenerate, but none of these phenomena would account for an {\it increase} in $1/\tau_1$ at low $T$. However, calculations of $1/\tau_1$ due to resonant carrier scattering in graphene do show this trend \cite{Kochan2014}, though the calculated $\tau_1$ is much larger and the temperature dependence is not as strong. Above this, in the range 100 to 300 K, the fast \slr\ rate is remarkably independent of temperature (\Cref{fig:HOPG_SLR_T1_Slow_Fast_Fraction}\textcolor{blue}{b}), reminiscent of the exchange narrowed high temperature regime of more dense spin systems \cite{Ding2013}. At even higher temperature, above 300 K, the slowing of the \slr\ rates and reduction in the fast fraction may be related to the onset of annealing as seen by \esr\ \cite{Kazumata1986}, consistent with a damage-related origin for the fast-relaxing component.

Below 50 K the \textit{amplitude} of the fast component also falls by nearly a factor of 2, indicating that there is a fraction of the \eli\ for which the fast relaxation appears thermally activated. There are thus three fractions: two are {\it constant} at about 20\%
(40\%) being slow- (fast-) relaxing up to 300 K. The remaining 40\% is slow relaxing at low temperature, but becomes fast by 50 K. We adopt a simple kinetic model with an activated transition from the slow to fast relaxing environment given by \Cref{eqn_sigmoid}, where $c$ represents the exponential prefactor, $k_B$ is the Boltzmann constant, and $E_a$ is the activation energy barrier. The total fast relaxing fraction $f_{\textrm{fast}}$ in the high and low temperature limits is defined by $\Gamma$ and $(-1 / \Delta) + \Gamma$, respectively. 

\begin{equation}
f_{\textrm{fast}}(T) = \Gamma - \frac{1}{\Delta + c \exp{(-E_a/k_B T})}
\label{eqn_sigmoid}
\end{equation}

\noindent From this, we obtain the sigmoidal curve shown in \Cref{fig:HOPG_SLR_T1_Slow_Fast_Fraction}\textcolor{blue}{c} and an $E_a$ of 18(2) meV. This may represent either an intrinsic energy scale, or it may be related to the Li-graphite defect. In the former case, it may result from a fraction of defects with thermally activated moments. Magnetic characterization of partially graphitized carbon reveals the coexistence of both stable and thermally activated magnetic moments \cite{Smith1980} that may correspond to the ionization of carbene-like defect sites \cite{Radovic2005} containing a lone pair that is nonmagnetic in the ground state, but becomes magnetic upon ionization. 

In the latter case, the fast component may correspond to \eli\ at the vacancy (the carbon substitutional site) where it is most strongly influenced by the local moment. A fraction of \eli\ that stops in the immediate vicinity of the defect may make a thermally activated site change, while another fraction is too far away (or is blocked from) making such a transition and remains slow relaxing. This kind of site change is known to occur around 150 K in Ag and Au \cite{2015-MacFarlane-SSNMR-68-1}. The layered structure of graphite may facilitate a lower temperature site change for \eli\ in the van der Waals gap adjacent to the vacancy. Emission channeling \cite{Wahl1997} could potentially test this hypothesis. In this case, the vacancy may act as a trap for the associated fraction of the \eli. At much higher fluence, there is some evidence that radiation damage traps Li on the graphene surface \cite{Nikko2016}. However, low mobility in the dilute limit does not require radiation damage, as demonstrated by the release measurements \cite{1989-Jungblut-PRB-40-10810}, and, in our experiments, neither the fast nor the slow relaxing components shows any evidence of mobility.    

\section{Conclusion \label{sec:conclusion}}

In summary, we have used $\beta$-detected \eli\ \nmr\ to explore the behavior of \ce{Li} in HOPG in the dilute limit. We obtain a broad resonance line with evidence for only a small average quadrupolar coupling constant of $\sim$2 kHz. We find the implanted Li is not appreciably mobile up to 400 K. This is surprising in the light of both experimental and computational results in the more concentrated regime. It is, however, in good agreement with dilute limit release measurements \cite{1989-Jungblut-PRB-40-10810}. It is not clear that it is consistent with high rate electrochemical deintercalation \cite{Dokko2010}, but it is unknown whether this procedure attains the thermodynamic $1'$ phase. Thus, we expose a key difference compared to the Li-dense stage compounds. This should motivate further development of theoretical understanding of the factors controlling Li mobility in the dilute limit. The fast \slr\ and temperature dependent \nmr\ shift suggest the dominant effect of a nearby local magnetic moment. This magnetic defect is intrinsic and likely related to implantation damage (\eg a carbon vacancy). Its behavior resembles neither an uncoupled (\ie Curie-like) nor Kondo defect, but may be explained with the resonant scattering picture developed for graphene.

\begin{acknowledgments}

We are grateful to R.\ Abasalti, D.\ Arseneau, S.\ Daviel, B.\ Hitti, B.\ Smith, and D.\ Vyas for their outstanding technical assistance during the \bnmr\ experiments. We thank F.\ McGee and D.\ Wang for their help with some of the early measurements. We have benefited from the advice and discussions with A.F.\ Hebard, K.\ Persson, J.R.\ Dahn, and K.\ Behnia. We also extend our gratitude to J.\ Dadap, H.\ Gautam, and S.\ Zhdanovich for their assistance with supplementary characterization measurements of the ZYA sample. This work was supported by NSERC Discovery grants provided to R.F.K.\ and W.A.M., NSERC CREATE IsoSiM fellowships provided to A.C.\ and R.M.L.M., and the SBQMI QuEST fellowships provided to D.F., V.L.K., and J.O.T.\ This work was, in part, supported by funding from the Max Planck-UBC-UTokyo Center for Quantum Materials and the Canada First Research Excellence Fund, Quantum Materials and Future Technologies Program.

\end{acknowledgments}

\appendix

\section{Effect of Annealing}
\label{apx:anneal}

\begin{figure}[H]
\centering
\includegraphics[height=2.5in]{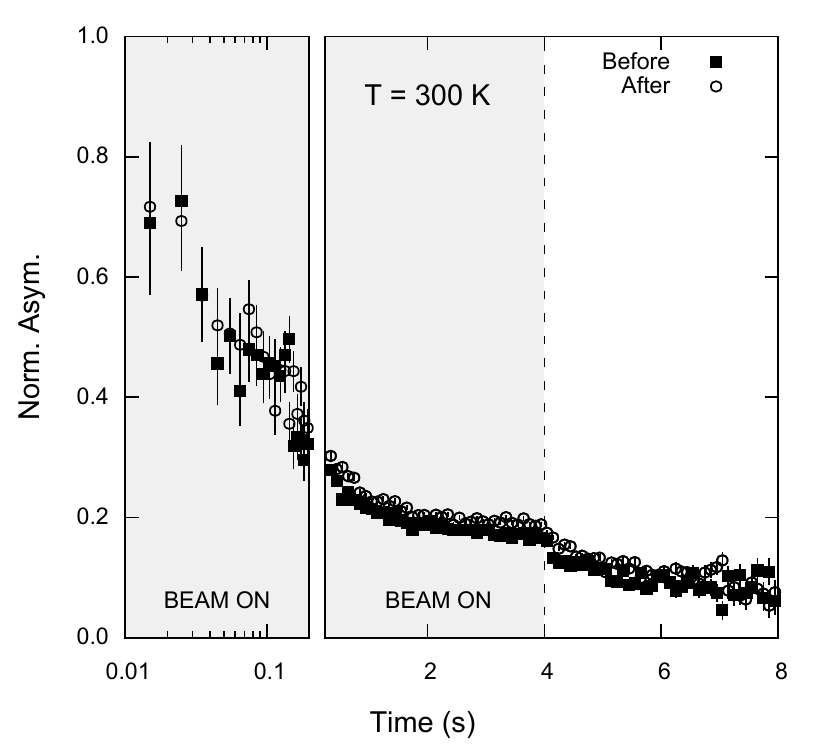}
\caption{Comparison of the \hopg\ \slr\ spectra at $B_0 =$ 6.55 T and 300 K, both before and after the annealing. The data here are plotted on a logarithmic scale at early times to make clear that the $1/T_1$ fast-relaxing components are very nearly identical. The data are unbinned in the logarithmic scale section and binned by a factor of 5 at later times. The data here are normalized to $A_0 = 0.095$, as noted in \Cref{subsec:SLR}.}
\label{fig:annealing_SLR}
\end{figure}

To test whether the ferromagnetism exhibited by some \hopg\ samples played any role in our \slr\ data, we adapted the high temperature annealing procedure from Ref. \onlinecite{2012-Miao-C-50-1614} and applied it to our ZYA sample. The peak temperature achieved during the 30 min annealing in turbo pumped high vacuum was $\sim$2400 $^{\circ}$C, slightly higher than Ref. \onlinecite{2012-Miao-C-50-1614} so as to compensate for poor thermal contact between our sample and its tantalum container. The SLR was measured before and after the annealing. The resulting data are compared in \Cref{fig:annealing_SLR}. From this, it is clear that the relaxation is unaffected by the annealing.

%% FIGURE %%

%% FIGURE %%

\section{Choice of SLR Fit Function}
\label{apx:SLR_fit_func}

One may consider whether or not the biexponential model is the optimal choice for fitting the SLR data. The interpretation of the data could be simplified by a stretched exponential function given by \Cref{eqn_str_exp} for a relaxation rate $\lambda = 1/T_1$ and a stretching exponent $\beta \in [0,1]$.

\begin{equation}
p_z(t) = \exp[-(\lambda t)^\beta]
\label{eqn_str_exp}
\end{equation}

\begin{figure}[H]
\centering
\includegraphics[height=2.5in]{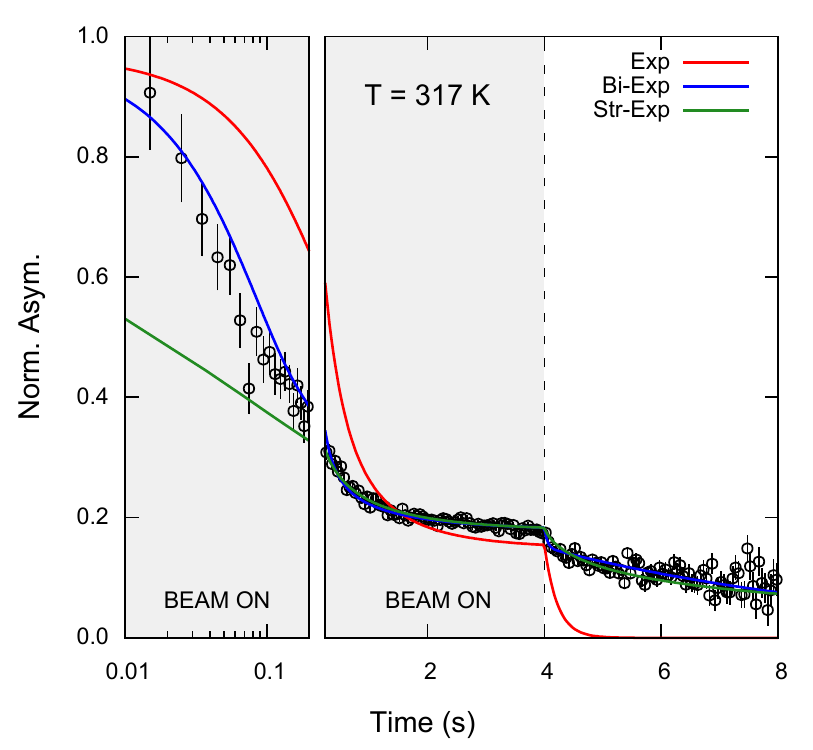}
\caption{Comparison of \slr\ fit types for $B_0 =$ 6.55 T and 317 K. To highlight the inadequacy of the single exponential and stretched exponential fits at early times, the data is presented on a logarithmic scale for $t\leq1$ s where $1/T_1^{\textrm{fast}}$ predominates. As before, the data and fits are normalized to $A_0 =0.095$.}
\label{fig:SLR_fit_type_compare}
\end{figure}

\noindent However, the stretched exponential cannot account for the relaxation at early times, as illustrated in \Cref{fig:SLR_fit_type_compare} on an expanded logarithmic time scale. An essential feature of the data is that the (statistical) uncertainty is highly inhomogeneous in time due to the \eli\ lifetime. The $\chi^2$ is thus heavily weighted by the statistically much more meaningful data in the vicinity of the trailing edge of the beam pulse where the count rates are maximal \cite{2022-MacFarlane-ZPC-236-757}. The three curves shown are the (weighted) least-squares fits for three relaxation models: single exponential, biexponential and stretched exponential. The stretched exponential consistently gives a very low exponent $\beta < 0.2$, which is typical when the data is instead biexponential. Single exponential fits are also poor and capture mainly the slow component. The next simplest function is the biexponential, but as noted in \Cref{subsec:SLR}, this introduces a fast relaxing component ($1/T_1^{\textrm{fast}}$) whose origin is not obvious \textit{a priori}. Its fraction $f_{\mathrm{fast}}$ is too large to be a background, and it is more than an order of magnitude faster than the slow \slr, such that the two components can be clearly distinguished by the fit. The single exponential fit overestimates $A(t)$ at early times, whereas it is underestimated by the stretched exponential. Interestingly, even when $\beta$ and $A_0$ are allowed to freely vary, the asymmetry at early times is still underestimated by the stretched fit.

%% FIGURE %%

%% FIGURE %%

\begin{figure}[H]
\centering
\includegraphics[height=2.5in]{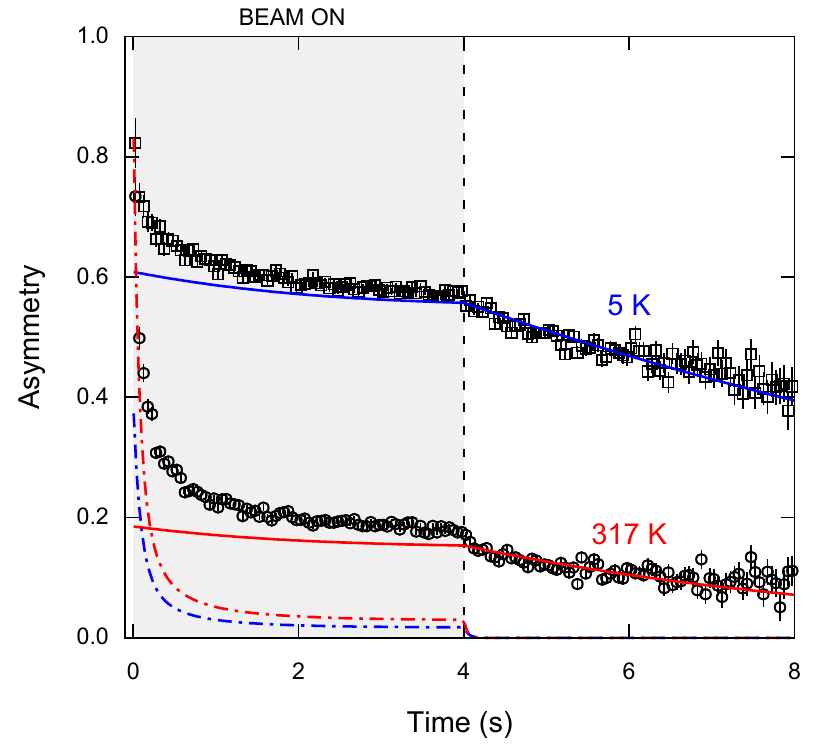}
\caption{Decomposition of the biexponential fits for the \slr\ spectra at 6.55 T. The slow and fast components of the relaxation are represented by solid and dashed lines, respectively. The spectra and fits are normalized to $A_0 =0.095$. At early times, the asymmetry is almost entirely captured in terms of the fast relaxation. At the trailing edge of the beam pulse and for times after, the slow relaxation predominates.}
\label{fig:SLR_fit_bi_exp_decomp}
\end{figure}

The phenomenology of the biexponential is clarified when it is deconstructed into its individual components, as shown in \Cref{fig:SLR_fit_bi_exp_decomp}, capturing well the fast relaxation at early times and the slower relaxation thereafter.

%% FIGURE %%

%% FIGURE %%

Although its origin is not obvious, we consider several possibilities. In cases where there are two crystallographically distinct lattice sites for the implanted \elip, the Li relaxation will differ. In the van der Waals gap, the na\"ive expectation is that the \elip\ will occupy many nearly equivalent sites, giving rise instead to a \textit{distribution} of relaxation rates probably giving a stretched exponential. The secondary source of relaxation is also unlikely to be caused by the \elip\ implanted in a macroscopic defect region (\eg grain boundaries or zig-zag edges), since this would 1) only account for a small \% of the sample space and 2) this would be expected to change with annealing.

Rather than an explanation owed to different environments sensed by the Li, the biexponential nevertheless implies distinct underlying mechanisms (\ie it appears to be an intrinsic feature of the dilute limit \eli\ \slr\ in all graphite samples studied), where their relative fraction varies weakly, except at the temperature extremes. On warming from 5 K, the weight of the slow component diminishes, possibly due to thermally activated local moment formation. The relaxation slows again above 320 K, though we find no evidence this is due to the Li becoming mobile. The persistent fast relaxation appears to be associated with some form of paramagnetism, which is discussed further in \Cref{sec:discussion}. Since the \eli\ probe is sensitive to both magnetic and electric quadrupolar effects, the origin of the slower relaxation may then be owed to the latter, and the extent of this may be tested by comparing the $^{8}$Li SLR rate to that of the heavier radioisotope $^{9}$Li (\eg see Ref. \onlinecite{2017-Chatzichristos-PRB-96-014307}).

\section{Helicity-Resolved Resonances}
\label{apx:split_hel}

The resonance spectra in \Cref{subsec:RES} combine the measured asymmetry of the two senses of the spin-polarization (\ie beam helicities). It is often useful, however, to compare the helicities separately to identify quadrupolar splitting, because opposite quadrupolar satellites should occur only in opposite helicities \cite{2015-MacFarlane-SSNMR-68-1}. 

%% FIGURE %%

\begin{figure}[H]
\centering
\includegraphics[height=2.5in]{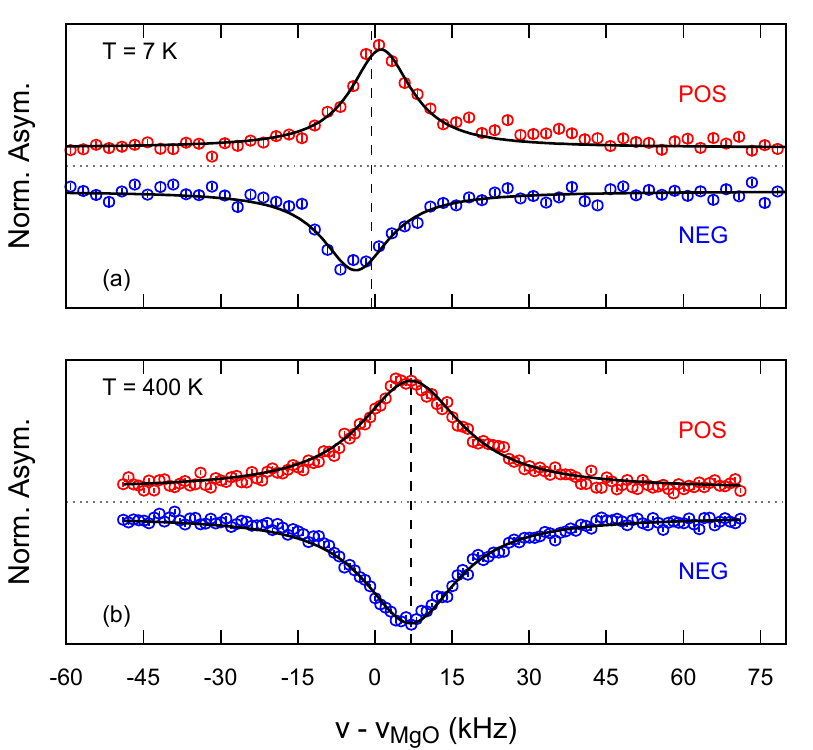}
\caption{The resolved helicity resonance lines at 7 K and 400 K for $B_0 =$ 6.55 T. The data are plot relative to the \ce{MgO} reference frequency $\nu_{\textrm{MgO}}$ (\Cref{apx:shift}) and the asymmetry is normalized to its baseline measurement. To assist the comparison, the asymmetry data are also vertically offset by a small \% centered about the horizontal dashed line. The vertical dashed line indicates the peak position of the combined resonance. The solid lines are fits to single Lorentzian functions.}
\label{fig:RES_split_hel}
\end{figure}

%% FIGURE %%

In the absence of resolved satellites, this may still be reflected in a difference between the two helicity spectra. Specifically, the resonance position will not be the same in the two helicities, and the difference is a measure of the average strength of this unresolved quadrupolar splitting. As shown in \Cref{fig:RES_split_hel}, at 7 K, peak positions differ by $\sim$4.9 kHz between the two helicities, suggestive of a weak quadrupolar coupling and consistent with the frequency comb measurements (see \Cref{fig:HOPG_RES_1w_Stacked}). In contrast, at 400 K, the helicity difference is unresolved. This is a result of both the broadening and increase in amplitude of the unsplit resonance line between 7-400 K, which masks the difference between the helicity-resolved peak positions at 400 K. 

\section{NMR Shift}
\label{apx:shift}

The Lorentzian fits provide the absolute line position as a function of temperature. With the magnet in persistence mode, the resonance frequency in an \ce{MgO} single crystal at 300 K provides a frequency reference ($\nu_{\textrm{MgO}}$) for the shift $\delta$ (in ppm) defined by \Cref{eqn_shift}.

\begin{equation}
\delta = \frac{\nu - \nu_{\textrm{MgO}}}{\nu_{\textrm{MgO}}} \times 10^6
\label{eqn_shift}
\end{equation}

\noindent The high temperature data between 300 and 400 K were taken in a separate run, so a separate calibration was necessary. $\nu_{\textrm{MgO}}$ is 41.271 692(6) MHz between 7 and 200 K, and 41.268 920(2) MHz at higher temperature. This `raw' resonance shift $\delta$ (plotted in \Cref{fig:HOPG_RES_width_shift}\textcolor{blue}{b}) is then corrected for the effect of demagnetization following \Cref{eqn_demag}, where $\chi_0$ is the (diamagnetic) dimensionless volumetric susceptibility of pure graphite (CGS units) and $N$ is the unitless demagnetizing factor determined by the shape of the sample \cite{1945-Osborn-PR-67-351}.

\begin{equation}
\delta_{\textrm{C}} = \delta + 4\pi \left[N-\frac{1}{3} \right] \chi_0(T)
\label{eqn_demag}
\end{equation}

We use an aggregate of $\chi_0$ measurements reported in Refs. \onlinecite{1939-Krishnan,1941-Ganguli,1998-Kaburagi} spanning the relevant temperature range. Note that because $\chi_0$ is large, demagnetization has a substantial effect.
For $B_0 = 6.55$ T oriented perpendicular to the sample face and the dimensions noted in \Cref{sec:experiment},
we estimate $N \approx$ 0.86. The resulting corrected values $\delta_{\textrm{C}}$ are shown in \Cref{fig:HOPG_RES_width_shift}\textcolor{blue}{c}, and the additive correction in \Cref{eqn_demag} is shown as the dashed line in \Cref{fig:HOPG_RES_width_shift}\textcolor{blue}{b}.

\clearpage

%% BIBLIOGRAPHY %%

\bibliography{manuscript.bib}

\end{document}

%% file: preamble/packages.tex
\usepackage[T1]{fontenc} 
\usepackage[utf8]{inputenc}
\usepackage{newtxtext}
\usepackage{newtxmath}
\usepackage{graphicx}
\usepackage{siunitx} 
\usepackage[version=4]{mhchem}
\usepackage{booktabs}
\usepackage[english]{babel}
\usepackage{csquotes}
\usepackage{microtype}
\usepackage[unicode,colorlinks=true,allcolors=blue]{hyperref}
\usepackage{cleveref}
\usepackage{float}
\usepackage{comment}
\usepackage{xcolor}
\usepackage{upgreek}

%% file: preamble/abbreviations.tex
\newcommand{\nmr}{NMR}

\newcommand{\bnmr}{$\beta$-NMR}

\newcommand{\slr}{SLR}
\newcommand{\efg}{EFG}

\newcommand{\rf}{RF}

\newcommand{\stm}{STM}
\newcommand{\esr}{ESR}

\newcommand{\eli}{\textsuperscript{8}{Li}}

\newcommand{\elip}{\textsuperscript{8}{Li}\textsuperscript{+}}

\newcommand{\lip}{{Li}\textsuperscript{+}}

\newcommand{\hopg}{HOPG}

\newcommand{\bpp}{BPP}
\newcommand{\stf}{SF}

\newcommand{\ie}{i.e., }
\newcommand{\eg}{e.g., }